\texsis
\overfullrule=0pt
\def\el{\overline \epsilon}
\def\mh{\hat{M}}
\def\ml{\overline M}
\def\ms{\buildrel \sim \over M}
\def\sb{\overline s}
\def\ss{\buildrel \sim \over S}
\def\ul#1{\underbar{#1}}
\def\xs{\buildrel \sim \over x}
\paper
\titlepage
\hbox{\space}\vskip 1in
\title Orbit Dynamics for Unstable Linear Motion
\endtitle
\author
George Parzen
March 29, 1996
BNL-63074
\endauthor
\abstract
A treatment is given of the orbit dynamics for linear unstable motion
that allows for the zeros in the beta function and makes no assumptions
about the realness of the betatron and phase functions.  The phase
shift per turn is shown to be related to the beta function and the
number of zeros the beta function goes through per turn.  The solutions
of the equations of motion are found in terms of the beta function.
\endabstract
\endtitlepage

\section{Introduction}

In the case of linear unstable motion, the beta function can be zero
at some points in the lattice.  
Because of the zeros in the beta function, and other assumptions often
made about the realness of the beta function and phase function, the
usual treatment given for stable motion does not carry over to the case
of unstable motion.  A treatment is given below, that allows for the
zeros in the beta functions and does not make assumptions about the
realness of the betatron and phase functions.

It will be shown that the solutions of the equations of motion can be
written in the form
$$\eqalign{
x &= \beta^{1\over 2} \exp(\pm\psi) \cr
\psi &= P \int_{s_0}^s {ds\over\beta} + i{\pi\over 2} N_z \cr}
\eqno{\hbox{(1-1)}}$$
$N_z$ is the number of times $\beta(s)$ goes through zero between
$s_0$ and $s$.  $P$ indicates the principle value of the integral.  
The solutions of the equations of motion can also be written as
$$x=\exp [\pm \mu s/L] f(s) \eqno{\hbox{(1-2)}}$$
where $f(s)$ is periodic and $L$ is the length of one turn.  It will
be shown that for unstable motion
$$\eqalign{
\mu &= 2\pi (g+i q/2) \cr
q &= {1\over 2} N_z \cr
g &= {P\over 2\pi} \int_0^L {ds\over\beta} \cr} \eqno{\hbox{(1-3)}}$$
where $N_z$ is the number of zeros the beta function goes through in one
turn.  $P$ indicates the principle value of the integral.

Often, the case of unstable linear motion is found when a gradient
perturbation is applied to a lattice whose unperturbed $\nu$-value
is close to $q/2$, $q$ being some integer.  In this case, perturbation
theory will show [1] that the solutions have the form given by Eq. (1-1)
where $q/2$ is the half integer close to the unperturbed to the
$\nu$-value.  In the general case, where the unstable motion cannot be
viewed as due to a perturbing gradient then the value of $q$ is given
by ${1\over 2} N_z$ where $N_z$ is the number of zeros in the beta function
in one turn.

It will also be shown that near a zero of the beta function at $s=s_1$,
$\psi$ will become infinite and the dominant term is $\psi$ is given by
$$\psi \sim \pm {1\over 2} \log (s-s_1) \eqno{\hbox{(1-4)}}$$

\section{The Definition of the Beta Function}

The linear parameters can be defined in terms of the elements of the
one period transfer matrix.  The $2\times 2$ transfer matrix, $M$, is
defined by
$$\eqalign{
x(s) &= M(s,s_0) x(s_0) \cr
x &= \pmatrix{ x \cr p_x \cr} \cr} \eqno{\hbox{(2-1)}}$$
The one period transfer matrix is defined by
$$\mh(s) = M(s+L,s) \eqno{\hbox{(2-2)}}$$
where the lattice is assumed to be periodic with the period $L$.  The 
matrix $M$ is assumed to be symplectic
$$\eqalign{
M\ml &= I \cr
\ml &= \ss \ms S \cr
S &= \pmatrix{0 & 1 \cr -1 & 0 \cr} \ \ I = \pmatrix{ 1 & 0 \cr
0 & 1 \cr} \cr} \eqno{\hbox{(2-3)}}$$
$\ss$ is the transpose of $S$.  Also $|M|=1$ where $|M|$ is the
determinant of $M$.  One can show that $\mh(s)$ and $\mh(s_0)$
are related by
$$\mh(s) = M(s,s_0) \mh(s_0) M(s_0,s) \eqno{\hbox{(2-4)}}$$
It follows from Eq. (2-4) that $\mh_{11}+\mh_{22}$, the trace of $\mh$,
is independent of $s$.  For unstable motion it is assumed that
$|\mh_{11}+\mh_{22}|>2$.  This may be shown to lead to unstable
exponentially growing motion.

One can now introduce the constant parameter $\mu$ defined by
$$\cosh \mu = {1\over 2} \left(\mh_{11}+\mh_{22}\right) \eqno{\hbox{(2-5)}}$$
If $\mh_{11}+\mh_{22}$ is positive, then $\mu$ will be real.  However
if $\mh_{11}+\mh_{22}$ is negative then $\mu$ has to have the imaginary
part $iq\pi$ where $q$ is an odd integer.  In general, one can write
$$\eqalign{
\mu &= \mu_R+iq\pi \cr
\cosh \mu_R &= {1\over 2} \left| \mh_{11}+\mh_{22} \right| \cr}
\eqno{\hbox{(2-6)}}$$
where $q$ is an even integer if $\mh_{11}+\mh_{22}$ is positive, and
$q$ is an odd integer when $\mh_{11}+\mh_{22}$ is negative.  It will
be seen below that $\mu_R$ is the exponential growth per period;
that is, the growth per period for the unstable solution is $\exp (\mu_R)$.
It will also be seen below that $q$ is related to the number of zeros
in the beta function, $\beta(s)$, in a period, which is $2q$.

$\mu$ is related to the eigenvalues of $\mh$, $\lambda_1$ and $\lambda_2$,
where $\lambda_1+\lambda_2=\mh_{11}+\mh_{22}$ and $\lambda_1\lambda_2=1$ 
from $|\mh-\lambda I|=0$. It follows from Eq. (2-5) that
$$\eqalign{
\lambda_1 &= \exp (\mu) \cr
\lambda_2 &= \exp (-\mu) \cr} \eqno{\hbox{(2-7)}}$$

One can define the linear parameters, $\beta$, $\alpha$, $\gamma$, using
the elements of the one period transfer matrix.  If one uses the form of
the transfer matrix often used [2] for stable motion the linear parameters
will be imaginary for unstable motion.  To make the linear parameters
real, they will be defined here in terms of the one period transfer
matrix as
$$\mh = \left[ \matrix{ \cosh \mu+\alpha \sinh \mu & \beta \sinh \mu \cr
\gamma \sinh \mu & \cosh \mu-\alpha \sinh \mu \cr} \right] 
\eqno{\hbox{(2-8)}}$$
$$\beta\gamma = 1 - \alpha^2$$
$\beta$, $\alpha$, $\gamma$ are then given in terms of $\mh_{ij}$ as
$$\eqalign{
\beta &= (-1)^q \mh_{12}/\sinh \mu_R \cr
\alpha &= (-1)^q  (\mh_{11}-\mh_{22})/2\sinh \mu_R \cr
\gamma &= (1-\alpha^2)/\beta \cr} \eqno{\hbox{(2-9)}}$$
Eq. (2-6) does not specify the sign of $\mu_R$.  One can define the
sign of $\mu_R$ to be always positive.  Then $\beta$, $\alpha$, $\gamma$
can then be computed from the $\mh_{ij}$ using Eq. (2-9).  It will be
seen later that the sign of $\beta(s)$ can change within a period,
and $\beta(s)$ can be zero at certain values of $s$ for unstable motion.

Having defined the linear parameters, one can now find the relationships
among them, their connection with the growth rate, the emittance and the
solutions of the equations of motion.  The treatment usually given for
stable motion does not carry over for unstable motion, because if
often assumes that $\beta$ and $\psi$, as defined for stable motion,
are real and that $\beta$ is never zero.  One needs a treatment which
does not make assumptions about the realness of $\beta$, $\psi$, and allows
$\beta$ to go through zero.  This is given below.

\subsection{Differential Equations for $\beta$, $\alpha$, $\gamma$}

It is assumed that the linearized equations of motion can be written as
$$\eqalign{
{dx\over ds} &= A_{11} x + A_{12} p_x \cr
{dp_x\over ds} &= A_{21} x + A_{22} p_x \cr
A_{11} + A_{22} &= 0 \cr} \eqno{\hbox{(2-10)}}$$
In the large accelerator approximation, $A_{11}=A_{22}=0$ and
$A_{12}=1$.  We note that
$$\eqalign{
{d\over ds} M(s,s_0) &= A\ M (s,s_0) \cr
{d\over ds} M(s_0,s) &= -M (s_0,s) A \cr} \eqno{\hbox{(2-11)}}$$
The last equation follows from $M(s,s_0) M(s_0,s)=I$.  Then using
Eq. (2-4)
$$\mh(s) = M(s,s_0) \mh(s_0) \mh(s_0,s) \eqno{\hbox{(2-12)}}$$
one finds
$${d\mh\over ds} = A\ \mh - \mh \ A . \eqno{\hbox{(2-13)}}$$
$A$ is the $2\times 2$ matrix whose elements are the $A_{ij}$ of
Eq. (2-10).  Replacing $\mh$, using Eq. (2-8), in Eq. (2-13) gives
the result
$$\eqalign{
{d\beta\over ds} &= 2 A_{11} \beta - 2 A_{12} \alpha \cr
{d\alpha\over ds} &= -A_{21} \beta + A_{12} \gamma \cr
{d\gamma\over ds} &= 2 A_{21} \alpha - 2 A_{11} \gamma \cr}
\eqno{\hbox{(2-14)}}$$
The first equation in Eqs. (2-14) gives the connection between $\alpha$
and $\beta$
$$\alpha = {1\over A_{12}} \left( - {1\over 2} {d\beta\over ds} + A_{11}
\beta \right) \eqno{\hbox{(2-15)}}$$

\subsection{Differential Equation for $\beta$}

In this section, the differential equation for $\beta$ will be obtained
without making any assumptions about the form of the solutions of the
equations of motion.  For the sake of simplicity, the derivation will be
given for the large accelerator case which assumes $A_{11}=A_{22}=0$ and
$A_{12}=1$.

Introducing $b$, where $\beta=b^2$, Eqs. (2-14) can be written as
$$\eqalign{
b{db\over ds} &= - \alpha \cr
{d\alpha\over ds} &= K b^2+\gamma \cr
{d\gamma\over ds} &= - 2 K \alpha \cr
K &= -A_{21} \cr} \eqno{\hbox{(2-16)}}$$
The first two equations in (2-16) then give
$$\eqalign{
{d\over ds} (b{db\over ds}) &= -K b^2 - \gamma \cr
&= -K b^2 - \left[ {1\over b^2} (1-b^2({db\over ds})^2) \right] \cr
&= -K b^2 - {1\over b^2} + ({db\over ds})^2 \cr} \eqno{\hbox{(2-17)}}$$
one then gets
$$\eqalign{
{d^2b\over ds^2} + K b + {1\over b^3} &= 0 \cr
b &= \beta^{1\over 2} \cr} \eqno{\hbox{(2-18)}}$$
Eq. (2-8) differs frm the usual result for stable motion only in the sign
of the $1/b^3$ term.

\section{$\beta(s)$ for Unstable Motion from Perturbation Theory}

Before proceeding further in finding the solutions of equations of motion
for unstable motion, and their connection with the beta function,
$\beta(s)$, it will be helpful to examine a result for the beta function
found using perturbation theory.  In reference [1], the case was studied
where a particle doing stable motion with the unperturbed tune $\nu_0$ is
perturbed by a small gradient perturbation which opens up an unstable
stopband around $\nu_0=q/2$, where $q$ is some integer.

In reference [1], the two solutions of the equations of motion inside the
stopband are found using perturbation theory, and will be denoted here as
$x_1$ and $x_2$.  Using $x_1$ and $x_2$ one can find the components of
$\mh$, and in particular
$$\eqalign{
\mh_{12} &= {1\over W} \left\{ - x_1(s) x_2(s_0)+x_2(s)x_1(s_0) \right\} \cr
W &= x_1 {dx_2\over ds} - x_2 {dx_1\over ds} \cr} \eqno{\hbox{(3-1)}}$$
In the following, the large accelerator approximation is being used,
$A_{12}=1$, $A_{11}=A_{22}=0$.  $W$ corresponds to the Wronskian 
and is a constant of the motion and can be evaluated at any value of $s$.
One can find $\beta$ from $\mh_{12}$ evaluated at $s=s_0+L$, and Eq.
(2-9),
$$\beta = (-1)^q \mh_{12}/\sinh \mu_R \eqno{\hbox{(3-2)}}$$
The result for $x_1$ to lowest order [1], is
$$\eqalign{
x_1 &= \beta_0^{1\over 2}(s) \exp (g\theta) \cos(q\theta/2-(\delta_1+
\delta_2)/2) \cr
g &= [|\Delta\nu|^2 - (q/2-\nu_0)^2]^{1\over 2} \cr
\Delta\nu &= {1\over 4\pi} \int ds \beta_0(s) {G\over B\rho} \exp
(-iq\theta) \cr
d\theta &= ds/\nu_0\beta_0, \ \ \delta_1 = \hbox{ph} (\Delta\nu) \cr
\delta_2 &= \hbox{ph} [(q/2-\nu_0)+ig] \cr} \eqno{\hbox{(3-3a)}}$$
The gradient perturbation is $\Delta B_y=-G(s)x$.  $g$ is positive.
The result given by Eq. (3-3) has an error which is first order in the
perturbation.  The $x_2$ solution is obtained from Eq. (3-3) by
replacing $g$ by $-g$ and $\delta_2$ by $-\delta_2$.  $\beta_0$,
$\nu_0$ are the unperturbed beta function and tune.

One may note that $x_1$ can also be written as
$$x_1 = \beta_0^{1\over 2} \exp [(q+iq/2)\theta] [1+\exp[i(q\theta-
(\delta_1+\delta_2))]] \eqno{\hbox{(3-3b)}}$$
which makes more evident the floquet form of the solution.  One sees
that $\mu$, the phase change in one turn is given by
$$\mu=2\pi g + iq\pi$$
One can write $x_1$ and $x_2$ as
$$\eqalign{
x_1 &= \beta_0^{1\over 2} \exp (g\theta) h_1(\theta) \cr
x_2 &= \beta_0^{1\over 2} \exp (-g\theta) h_2(\theta) \cr
h_1(\theta) &= \cos (q\theta/2-(\delta_1+\delta_2)/2) \cr
h_2(\theta) &= \cos (q\theta/2-(\delta_1-\delta_2)/2) \cr} 
\eqno{\hbox{(3-4)}}$$
One then finds
$$\eqalign{
{dx_2\over d\theta} &= \beta_0^{1\over 2} \exp (-g\theta) h_2
[-g+h'_2/h_2+{1\over 2}\beta'_0/\beta_0] \cr
{x_1 dx_2\over d\theta} &= \beta_0 h_1 h_2 [-g+h'_2/h_2+{1\over 2}
\beta'_0/\beta_0] \cr
W &= {1\over \nu_0} [-2g h_1h_2+h_1h'_2 - h'_1 h_2] \cr
W &= {1\over \nu_0} \left[-2g \cos(q\theta/2-(\delta_1+\delta_2)/2)
\cos(q\theta/2 - (\delta_1-\delta_2)/2) \right. \cr
&- (q/2) \cos (q\theta/2-(\delta_1+\delta_2)/2) \sin(q\theta_2-
(\delta_1-\delta_2)/2) \cr
&+ \left. q/2 \cos (q\theta/2-(\delta_1-\delta_2)/2) \sin (q\theta/2 -
(\delta_1+\delta_2)/2)\right] \cr
W &= {-q\over 2\nu_0} \sin \delta_2 , \ \ \sin \delta_2 = g/|\Delta\nu| \cr}
\eqno{\hbox{(3-5)}}$$
where the $2gh_1h_2$ term in $W$ has dropped as being of high order
than the remaining term.

One finds for $\mh_{12}$ from Eq. (3-1)
$$\mh_{12} = {1\over W} (\beta_0(\theta) \beta_0(\theta_0))^{1\over 2}
\left[ - \exp [g(\theta-\theta_0)] h_1(\theta) h_2(\theta_0) 
+ \exp [-g(\theta-\theta_0)] h_1(\theta_0)h_2(\theta) \right] 
\eqno{\hbox{(3-6)}}$$
Putting $\theta=\theta_0+2\pi$ and noting that $h_1(\theta)=(-1)^q$ 
$h_1(\theta_0)$, $h_2(\theta)=(-1)^q$ $h_2(\theta_0)$, and $2\pi g =
\mu_R$, one gets
$$\mh_{12} = {-\beta_0\over W}(-1)^q 2 \sinh \mu_R \cos (q\theta_0/2-
(\delta_1+\delta_2)/2) \cos (q\theta_0/2-(\delta_1-\delta_2)) 
\eqno{\hbox{(3-7)}}$$
$\beta$ can then be found using Eqs. (3-2), (3-5) and (3-6)
$$\beta(\theta) = \beta_0(\theta) {4\nu_0\over q} {|\Delta\nu|\over g}
\cos (q\theta/2-(\delta_1+\delta_2)/2) \cos (q\theta/2 - (\delta_1-
\delta_2)/2) \eqno{\hbox{(3-8a)}}$$
Eq. (3-8a) can also be written as
$$\eqalign{
\beta(\theta) &= \beta_0(\theta) {2\nu_0\over q} {|\Delta\nu|\over g}
[\cos \delta_2 + \cos (q\theta-\delta_1)] \cr
g &= [|\Delta\nu|^2-(q/2-\nu_0)^2]^{1\over 2} \cr} \eqno{\hbox{(3-8b)}}$$

Eqs. (3-8) show that as a function of $\theta$, $\beta$ will change sign
and go through zero twice in each interval of $2\pi/q$.  These two
zeros are located at
$$q\theta = \delta_1\pm\delta_2+\pi \eqno{\hbox{(3-9)}}$$
As a function of $\nu_0$, $\beta$ becomes infinite near the edge of the
stopband $|q/2-\nu_0|=|\Delta\nu|$ and drops to a value of the order of
$\beta_0$ near the center of the stopband, $\nu_0=q/2$.  In one turn,
$\Delta\theta=2\pi$, $\beta$ has $2q$ zeros.  One sees that the number
of zeros of $\beta$ in one turn is related to the imaginary part of
$\mu$, which is $q\pi$.

This result that connects the number of zeros in $\beta$ in one turn
with the imaginary part of the change in the betatron phase in one turn,
was found here using perturbation theory, but will be shown to be generally
valid in the next section.

\section{Solutions of the Equations of Motion and the Beta Function}

For stable motion, the role of the beta function in the solutions of
the equations of motion is well known.  A similar result will be found
here for unstable motion.  The treatment usually given for stable
motion, does not carry over to unstable motion because of the assumptions
usually made about the realness of the betatron and phase functions, and
the absence of zeros in the beta function.

Let us write the solutions of the equations of motion as
$$\eqalign{
x &= b \exp (\psi) \cr
b &= \beta^{1\over 2} \cr} \eqno{\hbox{(4-1)}}$$
where $\beta$ and $b$ have been defined by Eq. (2-8).  Then $b$ has been
shown to obey, see Eq. (2-18),
$$\eqalign{
{d^2b\over ds^2} &+ K b + {1\over b^3} = 0 \cr
K &= -A_{21} \cr} \eqno{\hbox{(4-2)}}$$
The treatment given in this section is for large accelerator case which
assumes $A_{11}=A_{22}=0$ and $A_{21}=1$.  Similar results can be found
for the general case.  $x$ then obeys the equations
$${d^2x\over ds^2} + K x = 0 \eqno{\hbox{(4-3)}}$$
Putting the form of $x$ assumed in Eq. (4-1) into Eq. (4-3), and using
Eq. (4-2) for $b$ one gets
$${d^2\psi\over ds^2} + {2\over b} {db\over ds} {d\psi\over ds} +
\left({d\psi\over ds}\right)^2 - {1\over b^4} = 0 \eqno{\hbox{(4-4)}}$$
Putting $f=d\psi/ds$ one gets
$${df\over ds} + {2\over b} {db\over ds} f + f^2 - {1\over b^4} = 0
\eqno{\hbox{(4-5)}}$$
The solutions of Eq. (4-5) are
$$f = \pm (1/b^2) = \pm 1/\beta, \eqno{\hbox{(4-6)}}$$
Thus
$$\psi = \pm \int_{s_0}^s {ds\over\beta} \eqno{\hbox{(4-7)}}$$
and the two solutions of the equations of motion are
$$x = \beta^{1\over 2} \exp \left(\pm \int_{s_0}^s {ds\over\beta}\right)
\eqno{\hbox{(4-8)}}$$

One may note that in deriving Eq. (2-8) no assumption was made about the
realness of $\beta$ or $\psi$.  However, there is a problem with the
result for unstable motion, as in the case of unstable motion $\beta(s)$
will go through zero.  To evaluate the integral when $\beta(s)$ has zeros,
Eq. (4-2) will be replaced by
$$\psi = \lim_{\epsilon\rightarrow 0} \int_{s_0}^s {ds\over\beta-
i\epsilon} \eqno{\hbox{(4-9)}}$$
where $\epsilon$ is a positive small quantity.  It can be shown that
Eq. (4-9) gives (see section 6)
$$\psi = P \int_{s_0}^{ds} {ds\over\beta} + \sum_{s_n} {i\pi\over
|\beta'(s_n)|} \eqno{\hbox{(4-10)}}$$
where $s_n$ are the locations of the zeros of $\beta(s)$ from $s_0$ to 
$s$.  $P$ represents the principle part of the integral.

One can also show that $\beta'(s)=\pm 2$ at the zeros of $\beta(s)$.
Since $\beta\gamma=\alpha^2-1$, then $\alpha=\pm 1$ when $\beta=0$.
Since $\beta'=-2\alpha$, $\beta'=\mp 2$ when $\beta=0$.  One can now
write Eq. (4-10) as
$$\psi = P \int_{s_0}^s {ds\over\beta} + {i\pi\over 2} N_z
\eqno{\hbox{(4-11)}}$$
where $N_z$ is the number zeros in $\beta(s)$ in $s_0$ to $s$.

One may notice that the imaginary part of $\psi$ has on an unusual
dependence on $s$.  It is constant in between zeros of $\beta(s)$ and
jumps by $\pi/2$ at each zero of $\beta(s)$.  One can use Eq. (4-11)
to find the change in $\psi$ over one turn, $\psi(s+L)-\psi(s)$, and find
$$\psi(s+L)-\psi(s) = P \int_{s}^{s+L} {ds\over\beta} + 
iq\pi \eqno{\hbox{(4-12)}}$$
where $2q$ is the number of zeros in $\beta(s)$ in one turn, and $L$
is the length of one turn.  For simplicity, it is being assumed that
the period $L$ is one turn.  Since $\beta(s)$ is a periodic function,
the number of zeros of $\beta(s)$ in one turn has to be even.  If one
defines the tune as the imaginary part of $\psi(s+L)-\psi(s)$ divided
by $2\pi$, then one has
$$\hbox{tune} = q/2 . \eqno{\hbox{(4-13)}}$$
Eq. (4-13) shows the connection between the tune and the number of zeros
in the beta function in one turn.  The real part of $\psi(s_0+L)-
\psi(s_0)$ gives the exponential growth in one turn.  If one defines
the exponential growth factor, $g$, to be the real part of $\psi(s+L)-
\psi(s)$ divided by $2\pi$
$$g = {P\over 2\pi} \int_s^{s+L} {ds\over\beta} \eqno{\hbox{(4-14)}}$$

Another apparent difficulty with the solutions given by Eq. (4-8) is that
at the $s$ value where $\beta(s)$ is zero, both solutions appear to go
to zero being proportional to $\beta^{1\over 2}$.  This is not possible
as the $x$ motion which is a linear combination of these two solutions
would then also have to go to zero at this $s$ value.  It will now be
shown that one of the solutions will not go to zero at the zeros of
$\beta(s)$.

Let $s_1$ be a zero of $\beta(s)$.  Then near $s=s_1$, $b=\beta^{1\over 2}$
goes to zero like $(s-s_1)^{1\over 2}$.  However, it is shown in section
6, that near $s=s_1$ that $\psi$ become infinite like $(1/\beta'(s_1))
\log (s-s_1)$.  Note that $\beta'(s_1) =\pm 2$, and that $b\exp(\psi)$
goes like $(s-s_1)^{1\over 2} (s-s_1)^{\pm{1\over 2}}$.  Depending on
the sign of $\beta'(s)$, $b\exp(\psi)$ may or may not go to zero at
$s=s_1$.  If $b\exp(\psi)$ does go to zero, then $b\exp(-\psi)$ will
not go to zero.  Thus one of the two solutions will not go to zero
at $s=s_1$.

It is interesting to note that the solutions given by Eq. (4-8) can be
chosen to be real.  Let us start at the $s$ value $s_0$ which is
assumed to be in a region where $\beta(s)$ is positive and let $s_1$
be the location of the first zero in $\beta(s)$ after $s=s_0$.  In the
region $s_0$ to $s_1$, the solution $b\exp(\psi)$ is real, as $\beta$ and
$\psi$ given by Eq. (4-9) are both real.  After $s=s_1$, $\beta$ becomes
negative and $b=\beta^{1\over 2}$ becomes pure imaginary.  However $\psi$
jumps at $s=s_1$ by $i\pi/2$.  Thus the solution $b\exp(\psi)$ remains
real just after $s=s_1$.  One can continue in this way through the
entire lattice with $\beta$ and $\psi$ changing suddenly after each
zero of $\beta(s)$ so as to keep the solutions real.  This result is
consistent with the result found in reference [1], that the eigenvalues
and eigenfunctions of the one period transfer matrix are real in a
linear half integer stopband.

\subsection{Eigenvectors of the Transfer Matrix}

The eigenvectors of $\mh$ will now be found in terms of $\beta$, $\alpha$
and $\psi$.  It will also be shown that the eigenvalues are given by
$\exp(\pm\Delta\psi)$, so that $\Delta\psi=\mu$, where $\Delta\psi=
\psi(s+L)-\psi(s)$.

Starting from $x=\beta^{1\over 2}\exp(\pm\psi)$, one can find the
corresponding $p_x$ from Eq. (2-10)
$$\eqalign{
p_x &= {1\over A_{12}} \left\{ {dx\over ds} - A_{11} x \right\} \cr
p_x &= \beta^{-{1\over 2}} (-\alpha\pm 1) \exp(\pm \psi) \cr
\alpha &= {1\over A_{12}} \left( - {1\over 2} {d\beta\over ds} +
A_{11}\beta\right) \cr} \eqno{\hbox{(4-15)}}$$
The two solutions can then be written as
$$\eqalign{
x_1 &= {\beta^{1\over 2} \brack \beta^{-{1\over 2}}(-\alpha+1)} \exp
[\psi], \ \ x_2 = {\beta^{1\over 2} \brack \beta^{-{1\over 2}} (-\alpha
-1)} \exp [-\psi] \cr
\psi &= \lim_{\epsilon\rightarrow 0} \int_0^s 
{ds\over\beta-i\epsilon} \cr} \eqno{\hbox{(4-16)}}$$
These two solutions are the eigenvectors of $\mh$ as
$$\eqalign{
\mh x_1 &= \exp (\Delta\psi) x_1 , \ \ \mh x_2 = \exp(-\Delta\psi) x_2 \cr
\Delta\psi &= \psi(s+L)-\psi(s) \cr} \eqno{\hbox{(4-17)}}$$
and the eigenvalues of $\mh$ are
$$\lambda_1 = \exp(\Delta\psi) \ \ \ \lambda_2 =\exp(-\Delta\psi)
\eqno{\hbox{(4-18)}}$$
Comparing Eq. (4-18) with Eq. (2-7), one sees that $\mu=\Delta\psi=
\psi(s+L)-\psi(s)$.

Since $x/\exp(\pm\mu s/L)$ is a periodic function, one can write $x$ as
$$\eqalign{
x &= \beta^{1\over 2} \exp [\pm\mu s/L] f(s) \cr
x &= \beta^{1\over 2} \exp [\pm 2\pi(g+iq/L)s/L] f(s) \cr}
\eqno{\hbox{(4-19)}}$$
where $f(s)$ is periodic with period $L$.  $q$ here is defined by $2q$
is the number of zeros in $\beta(s)$ in one turn.  One also has
$$\mu=2\pi(g+iq/2) \eqno{\hbox{(4-20)}}$$
where it is assumed that the period is one turn.

To summarize, it has been found that if $2q$ is the number zeros in the
beta function in one turn, then the eigenvalues of the one period 
transfer matrix, $\exp(\pm\mu)$ are given by
$$\eqalign{
\mu &= 2\pi g +iq\pi \cr
g &= {P\over 2\pi} \int_0^L {ds\over\beta} \cr} \eqno{\hbox{(4-21a)}}$$
where $P$ indicates the principle part of the integral, and the solutions
of the equations of motion are given by
$$\eqalign{
x &= \beta^{1\over 2} \exp(\pm\psi) \cr
\psi &= \lim_{\epsilon\rightarrow 0} \int_{s_0}^s 
{ds\over\beta-i\epsilon} \cr} \eqno{\hbox{(4-21b)}}$$

\section{The Emittance Invariant}

The emittance invariant can be found from the Lagrange invariant for
symplectic motion.  If $x_1$ and $x_2$ are two solutions of the
equation of motion then [2]
$$\xs_2 S x_1 = \hbox{constant} \eqno{\hbox{(5-1)}}$$
It is assumed that the lattice is periodic, so that the coefficients in
the linearized equations ofmotion are periodic in $s$ with the period
$L$.  Thus if $x(s)$ is a solution then $x(s+L)$ or $\mh(s) x(s)$ is 
also a solution.  In Eq. (3-1) putting $x_1=x$, $x_2=\mh x$ then one
gets the invariant [3]
$$\xs s \mh x = \hbox{constant} \eqno{\hbox{(5-2)}}$$
Using Eq. (2-6) for $\mh$ one finds
$$\xs s \mh x = -\sin\mu(-\gamma x^2+2\alpha x p_x + \beta p_x^2)
\eqno{\hbox{(5-3)}}$$
Thus Eq. (5-3) gives the emittance invariant
$$\eqalign{
\epsilon &= \gamma x^2 - 2\alpha x p_x - \beta p_x^2 \cr
\epsilon &= {1\over\beta} (x^2-(\alpha x + \beta p_x)^2) \cr
\beta\gamma &= 1 - \alpha^2 \cr} \eqno{\hbox{(5-4)}}$$
Eq. (5-4) shows that the curve $\epsilon=$ constant is a hyperbola.
In the case of stable motion, the curve $\epsilon=$ constant is an
ellipse and $\epsilon$ gives the phase space area enclosed by the
ellipse.  For unstable motion, $\epsilon$ does not have a simple
interpretation in terms of phase space, also $\epsilon$ can be negative.

Eq. (5-4) suggests introducing the new symplectic variable $\eta$,
$p_\eta$ where
$$\eqalign{
\pmatrix{ \eta \cr p_\eta \cr} &= G \pmatrix{x \cr p_x \cr} \cr
G &= \pmatrix{ \beta^{-{1\over 2}} & 0 \cr \alpha\beta^{-{1\over 2}}
& \beta^{1\over 2} \cr} \cr
\eta &= \beta^{-{1\over 2}} x, \ \ p_n = \beta^{-{1\over 2}} (\alpha x
+\beta p_x) \cr
|G| &= 1 \cr} \eqno{\hbox{(5-5)}}$$
The emittance invariant can then be written as
$$\epsilon = \eta^2 - p_\eta^2 \eqno{\hbox{(5-6)}}$$

\subsection{Minimum Amplitude}
Eq. (5-4) shows that the particle will move in a hyperbola.  Under
certain conditions, the particle will first move to smaller $x$ or
$p_x$ before the amplitude of the motion starts to grow exponentially.
It will be shown below that $x$ and $p_x$ can attain the minimum
$$\eqalign{
x_{\rm min} &= (\beta\epsilon)^{1\over 2} , \cr
p_{x,{\rm min}} &= (-\gamma\epsilon)^{1\over 2} , \ \ \gamma\beta=
1-\alpha^2 \cr} \eqno{\hbox{(5-7)}}$$
Eq. (5-7) shows that $x$ will have a mimimum when $\beta\epsilon>0$,
and $p_x$ will have a mimimum when $\beta\epsilon<0$ if $|\alpha|<1$
or $\beta\epsilon>0$ if $|\alpha|>1$.

The minimum given by Eq. (5-7) can be computed from Eq. (5-4) by
computing $d\epsilon/ds$ and putting $d\epsilon/ds=0$ and either
$dx/ds=0$ or $dp_x/ds=0$.

\subsection{Asymptotes and Rotation Angle}
If one plots $\beta p_x$ versus $x$, one can ask what are the directions
of the symptotes of the hyperbola.  If $\delta_1$ and $\delta_2$ are the
angles with the $x$ axis for these asymptotes, then they are given by
$$\eqalign{
\tan \delta_1 &= 1-\alpha \cr
\tan \delta_2 &= -1 - \alpha \cr} \eqno{\hbox{(5-8)}}$$
These results can be found by assuming the asymptotic expansion for
$\beta p_x$, $\beta p_x=\tan\delta\  x+c_0+c_{-1}x^{-1}...$ and putting
this into the equation of the hyperbola, Eq. (5-4).  Collecting all
the $x^2$ terms and putting the coefficient of $x^2=0$ gives Eq. (5-8).

If one plots $\beta p_x$ versus $x$, then one can ask through what angle
this coordinate system has to be rotated to make the hyperbola have its
normal form.  This rotation angle is given by
$$\tan 2\theta = {2\alpha\over\alpha^2-2} \eqno{\hbox{(5-9)}}$$

\section{Phase Function Results when $\beta$ has Zeros}

In this section, the result for the phase function, $\psi$, given by
Eq. (4-10) will be derived.  Also, the behavior of $\psi$ when $s$ is
near the zeros of $\beta(s)$ will be studied.

First, let us consider the case where
$$\psi = \lim_{\epsilon\rightarrow 0} \int_{s_0}^s {ds\over\beta-i\epsilon}
\eqno{\hbox{(6-1)}}$$
$\epsilon>0$, and one assumes there is only one zero for $\beta(s)$
at $s=s_1$ between $s=s_0$ to $s=s$.  Then, one can write
$$\psi = P \int_{s_0}^s {ds\over\beta} + \int_{s_1-\delta}^{s_1+\delta}
{ds\over\beta-i\epsilon} \eqno{\hbox{(6-2)}}$$
where $\delta\rightarrow 0$ but $\delta \gg\epsilon$.  $P$ stands for
the principle part of the integral.  Near $s_1$ one can write $\beta =
\beta'(s_1) (s-s_1)+...$ and find
$$\eqalign{
\int_{s_1-\delta}^{s_1+\delta} {ds\over\beta-i\epsilon} &=
\int_{s_1-\delta}^{s_1+\delta} {ds\over\beta'(s_1)(s-s_1)-i\epsilon} \cr
&= {1\over\beta'(s_1)} \int_{-\delta}^{\delta} d\sb {(\sb+i\el)\over
\sb^2+\el^2}, \ \ \sb=s-s_1, \ \ \el=\epsilon/\beta'(s_1) \cr
&= {1\over\beta'(s_1)} {i\el\over |\el|} \pi \cr
&= {1\over |\beta'(s_1)|} i\pi \cr} \eqno{\hbox{(6-3)}}$$
If there are many zeros between $s_0$ to $s$ at $s=s_n$ one then finds
$$\psi = P \int_{s_0}^s {ds\over\beta} + \sum_{s_n} {i\pi\over |\beta'
(s_n)|} \eqno{\hbox{(6-4)}}$$

Now, it will be shown that near a zero of $\beta(s)$, like $s=s_1$,
$\psi$ becomes infinite like
$$\psi \sim \pm {1\over 2} \log (s-s_1) \eqno{\hbox{(6-5)}}$$
The $\pm$ corresponds to the sign of $\beta'(s_1)$.  We write $\psi$ as
$$\eqalign{
\psi &= A + B \cr
A &= \int_{s_0}^s ds \left\{{1\over\beta-i\epsilon}-{1\over\beta'(s_1)
(s-s_1)-i\epsilon}\right\} \cr
B &= \int_{s_0}^s {1\over\beta'(s_1)(s-s_1)-i\epsilon} \cr}
\eqno{\hbox{(6-6)}}$$
where $s$ is assumed to be close to $s_1$ but $s>s_1$.

The integral of $A$ has no pole near $s=s_1$ and $A$ does not become
infinite at $s=s_1$.  $B$ can be written as
$$\eqalign{
B &= {1\over\beta'(s_1)} \int {ds\over (s-s_1)-i\epsilon} \cr
B &= {1\over\beta'(s_1)} \left\{ \log (s-s_1-i\epsilon) - \log
(s_0-s_1-i\epsilon) \right\} \cr
B &\sim \pm {1\over 2} \log (s-s_1) \cr} \eqno{\hbox{(6-7)}}$$
where, in the last result, only the dominant term that becomes infinite
at $s=s_1$, has been kept and the result $\beta'(s_1)=\pm2$ has been used.

Thus near $s=s_1$, $\psi$ becomes infinite and the dominant term is given
by
$$\psi\sim\pm{1\over 2} \log(s-s_1) \eqno{\hbox{(6-8)}}$$
where the $\pm$ is chosen to correspond to the sign of $\beta'(s_1)$.

\nosechead{References}

\enumerate

\itm G. Parzen, Particle motion inside and near a linear half--integer
stopband, BNL Report, BNL-62036 (1995).

\itm E.D. Courant and H.S. Snyder, Theory of the alternating gradient 
synchrotron, Ann. Phys. \ul{3}, 1 (1958).

\itm J.S. Bell, Hamiltonian Mechanics, CERN Accelerator school
proceedings, CERN 87-03 (1987); Rutherford Lab. Report AERE T/R 1114
(1953).

\endenumerate

\bye